\documentclass[prb,aps,showpacs,onecolumn]{revtex4}
\usepackage{amssymb}
\usepackage{amsfonts}
\usepackage{graphicx}
\usepackage{amsmath}
\usepackage{bm}

\input{tcilatex}
\begin{document}

\title{Rational solutions for the Riccati-Schr\"{o}dinger equations
associated to translationally shape invariant potentials}
\author{Y. Grandati and A. B\'{e}rard}
\affiliation{Institut de Physique, ICPMB, IF CNRS 2843, Universit\'{e} Paul Verlaine, 1
Bd Arago, 57078 Metz, Cedex 3, France}

\begin{abstract}
We develop a new approach to build the eigenfunctions of a translationally
shape-invariant potential. For this we show that their logarithmic
derivatives can be expressed as terminating continued fractions in an
appropriate variable. We give explicit formulas for all the eigenstates,
their specific form depending on the Barclay-Maxwell class to which the
considered potential belongs.
\end{abstract}

\maketitle

\section{Introduction}

In the framework of SUSY quantum mechanics, the concept of shape invariance
is a key feature of explicit exact solvability\cite%
{Cooper,Gendenshtein,Dutt,Carinena2}. Combining the hierarchy of SUSY
partner hamiltonians with the shape-invariance condition, it is possible,
when the SUSY is unbroken, to obtain the entire spectrum of a SIP in terms
of the functions characterizing the partners' correspondence. Moreover it
gives an access to the eigenfunctions via a generalization of the harmonic
creation-annihilation operators. Among all SIP those for which the partner's
parameters are related by a translation (TSIP) play a particular role.
Indeed they are so far the only ones for which we have closed-form
expressions for the superpotentials and then for the eigenfunctions.

In this paper we propose an alternate way to determine the eigenfunctions of
a TSIP. For this we construct analytically the logarithmic derivatives of
the eigenfunctions, which we call hereafter the Riccati-Schr\"{o}dinger (RS)
functions of the system. They are solutions of a particular type of Riccati
equations depending upon the energy as a parameter that we will call here
Riccati-Schr\"{o}dinger or RS equations. Using the finite-difference B\"{a}%
cklund algorithm\cite{Carinena,Ramos,Fernandez,Mielnik,Adler1,Adler2} we
obtain a terminating continued fraction expression for the RS functions in
terms of the superpotential (a result outlined in a different way and in an
incomplete form by Kazimierz\cite{Kazimierz}). Then we consider two
categories of potentials which can be reduced to an harmonic or isotonic
form by a change of variable satisfying a constant coefficient Riccati
equation. We show that their ground state RS function is a first degree
polynomial or a first degree Laurent polynomial (depending on the category
under consideration) in the new variable and that they are translationally
shape invariant, giving general simple algebraic formula for the energy
spectrum. Combining these results, we obtain exact rational expressions for
the RS functions in terms of the new variable, which permits to recover in a
simple way the eigenfunctions of the system. Finally we establish the
equivalence between these two categories and the two Barclay-Maxwell classes
of TSIP\cite{barclay1,Bhaduri} which shows that the above construction
applies in fact to the whole set of TSIP.

\section{B\"{a}cklund algorithm}

\subsection{Invariance group on the set of Riccati equations}

As established by Cari\~{n}ena et al.\cite{Carinena,Ramos}, the \bigskip
finite-difference B\"{a}cklund algorithm is a consequence of the invariance
of the set of Riccati equations under a subset of the group $\mathcal{G}$ of
smooth $SL(2,\mathbb{R})$-valued curves $Map(\mathbb{R},SL(2,\mathbb{R}))$.
For any element $A\in $ $\mathcal{G}$ characterized by the matrix:

\begin{equation}
A(x)=\left( 
\begin{array}{cc}
\alpha (x) & \beta (x) \\ 
\gamma (x) & \delta (x)%
\end{array}%
\right) ,\quad \det A(x)=\alpha (x)\delta (x)-\beta (x)\gamma (x)=1,
\label{matrice}
\end{equation}%
the action of $A$ on $Map(\mathbb{R},\overline{\mathbb{R}})$ is given by:

\begin{equation}
w(x)\overset{A}{\rightarrow }\widetilde{w}(x)=\frac{\alpha (x)w(x)+\beta (x)%
}{\gamma (x)w(x)+\delta (x)}=\frac{\alpha (x)}{\gamma (x)}-\frac{1}{\gamma
(x)}\frac{1}{\gamma (x)w(x)+\delta (x)}.  \label{transfo}
\end{equation}

If $A$ acts on a solution of the Riccati equation:

\begin{equation}
w^{\prime }(x)=a_{0}(x)+a_{1}(x)w(x)+a_{2}(x)w^{2}(x)  \label{edrg}
\end{equation}%
we obtain a solution of a new Riccati equation:

\begin{equation}
\widetilde{w}^{\prime }(x)=\widetilde{a}_{0}(x)+\widetilde{a}_{1}(x)%
\widetilde{w}(x)+\widetilde{a}_{2}(x)\widetilde{w}^{2}(x)  \label{edr2}
\end{equation}%
the coefficients of which being given by $\overrightarrow{u}(x)=\left( 
\begin{array}{c}
u_{2}(x) \\ 
u_{1}(x) \\ 
u_{0}(x)%
\end{array}%
\right) $:

\begin{equation}
\overrightarrow{\widetilde{a}}(x)=M(A)\overrightarrow{a}(x)+\overrightarrow{W%
}(x),  \label{transfocoeff1}
\end{equation}%
where:

\begin{equation}
M(A)=\left( 
\begin{array}{ccc}
\delta ^{2}(x) & -\gamma (x)\delta (x) & \gamma ^{2}(x) \\ 
-2\beta (x)\delta (x) & \alpha (x)\delta (x)+\beta (x)\gamma (x) & -2\alpha
(x)\gamma (x) \\ 
\beta ^{2}(x) & -\alpha (x)\beta (x) & \alpha ^{2}(x)%
\end{array}%
\right) ,
\end{equation}

\begin{equation}
\overrightarrow{W}(x)=\left( 
\begin{array}{c}
W(\gamma ,\delta ;x) \\ 
W(\delta ,\alpha ;x)+W(\beta ,\gamma ;x) \\ 
W(\alpha ,\beta ;x)%
\end{array}%
\right)
\end{equation}%
and $W(f,g;x)=f(x)g^{\prime }(x)-f^{\prime }(x)g(x)$ is the wronskian of $%
f(x)$ and $g(x)$ in $x$. As noted in \cite{Carinena}, Eq.(\ref{transfocoeff1}%
) defines an affine action of $\mathcal{G}$ on the set of general Riccati
equations.

\subsection{Riccati-Schr\"{o}dinger equations}

To a one-dimensional Schr\"{o}dinger equation ($\hbar =1,m=\frac{1}{2}$) for
a potential $V(x)$:

\begin{equation}
\psi ^{\prime \prime }(x)+(E-V(x))\psi (x)=0  \label{eds}
\end{equation}%
the transformation:

\begin{equation}
w(x)=-\frac{\psi ^{\prime }(x)}{\psi (x)}
\end{equation}%
associates a particular Riccati equation of the form:

\begin{equation}
-w^{\prime }(x)+w^{2}(x)=V(x)-E  \label{edr3}
\end{equation}%
which corresponds to Eq.(\ref{edrg}) with the coefficients $%
a_{0}(x)=E-V(x),\ a_{1}(x)=0\ $\ and $a_{2}(x)=1$.

We'll call such an equation a Riccati-Schr\"{o}dinger (RS) equation and $%
w(x) $ a RS function. Note that, up to a $i$ factor, the RS function
identifies with the quantum momentum function at energy $E$ in the Quantum
Hamilton-Jacobi formalism (QHJ) of Leacock and Padgett\cite{bhalla,bhalla2}
From the knowledge of the RS function, we recover immediately the
corresponding wave function via:

\begin{equation}
\psi (x)\sim \exp \left( -\int^{x}w(s)ds\right)
\end{equation}%
and the Schr\"{o}dinger equation Eq(\ref{eds}) can be rewritten:

\begin{equation}
H\psi (x)=\left( L^{+}L+E\right) \psi (x)=E\psi (x)  \label{hamilton}
\end{equation}%
with

\begin{equation}
L=\frac{d}{dx}+w(x).
\end{equation}

To each node of $\psi (x)$ (which is necessarily simple) corresponds a
simple pole of $w(x).$ Moreover $w(x)$ decreases in the interval $\left[
x_{1},x_{2}\right] $, $x_{1}$ and $x_{2}$ being the turning points of the
classical motion.

\subsection{Finite difference B\"{a}cklund algorithm}

When applied to the RS equation Eq.(\ref{edr3}), Eq(\ref{transfocoeff1})
gives:

\begin{equation}
\left\{ 
\begin{array}{c}
\widetilde{a}_{2}(x)=\delta ^{2}(x)+\left( E-V(x)\right) \gamma
^{2}(x)+W(\gamma ,\delta ;x) \\ 
\\ 
\widetilde{a}_{1}(x)=-2\beta (x)\delta (x)-2\alpha (x)\gamma (x)\left(
E-V(x)\right) +W(\delta ,\alpha ;x)+W(\beta ,\gamma ;x) \\ 
\\ 
\widetilde{a}_{0}(x)=\beta ^{2}(x)+\alpha ^{2}(x)\left( E-V(x)\right)
+W(\alpha ,\beta ;x).%
\end{array}%
\right.  \label{transfocoeff3}
\end{equation}

The most general elements of $\mathcal{G}$ preserving the subset of RS
equations has been determined in \cite{Carinena}. Among them we find in
particular the elements of the form:

\begin{equation}
A(x)=\frac{1}{\sqrt{\lambda }}\left( 
\begin{array}{cc}
\phi (x) & \lambda -\phi ^{2}(x) \\ 
-1 & \phi (x)%
\end{array}%
\right) ,\ \lambda >0  \label{transfo2}
\end{equation}%
where $\phi (x)$ satisfies an RS equation with the same potential as Eq.(\ref%
{edr3}) but with a shifted energy:

\begin{equation}
-\phi ^{\prime }(x)+\phi ^{2}(x)=V(x)-\left( E-\lambda \right)
\end{equation}

With this choice Eq(\ref{transfocoeff3}) becomes%
\begin{equation}
\left\{ 
\begin{array}{c}
\widetilde{a}_{2}(x)=\frac{1}{\lambda }\left( -\phi ^{\prime }(x)+\phi
^{2}(x)+E-V(x)\right) =1 \\ 
\\ 
\widetilde{a}_{1}(x)=\frac{2\phi (x)}{\lambda }\left( -\phi ^{\prime
}(x)+\phi ^{2}(x)+E-V(x)\right) -2\phi (x)=0 \\ 
\\ 
\widetilde{a}_{0}(x)=\frac{\phi ^{2}(x)}{\lambda }\left( -\phi ^{\prime
}(x)+\phi ^{2}(x)+E-V(x)\right) +\lambda -2\phi ^{2}(x)-\phi ^{\prime
}(x)=V(x)+2\phi ^{\prime }(x)%
\end{array}%
\right.
\end{equation}%
and $\widetilde{w}(x)$ satisfies the RS equation:

\begin{equation}
-\widetilde{w}^{\prime }(x)+\widetilde{w}^{2}(x)=\widetilde{V}_{\phi
}(x)-\lambda
\end{equation}%
where $\widetilde{V}_{\phi }(x)=V(x)+2\phi ^{\prime }(x)$.

Suppose then that the spectrum of the hamiltonian $H$ associated to $V(x)$
is $\left( E_{l},\psi _{l}(x)\right) $ with $l\geq 0$ not necessarily
discrete. The corresponding RS functions $w_{l}(x)=-\psi _{l}^{\prime
}(x)/\psi _{l}(x)$ satisfy the RS equations associated with the different
values of the energy:

\begin{equation}
-w_{l}^{\prime }(x)+w_{l}^{2}(x)=V(x)-E_{l}.  \label{edr4}
\end{equation}

Starting from a given $w_{k}(x)$, for every value of $l$ such that $%
E_{l}>E_{k}$ , we can build a element $A_{kl}\in \mathcal{G}$ of the form:

\begin{equation}
A_{kl}(x)=\frac{1}{\sqrt{E_{l}-E_{k}}}\left( 
\begin{array}{cc}
w_{k}(x) & E_{l}-E_{k}-w_{k}{}^{2}(x) \\ 
-1 & w_{k}(x)%
\end{array}%
\right)  \label{transfoback}
\end{equation}%
which transforms $w_{l}$ as:

\begin{equation}
w_{l}(x)\overset{A_{kl}}{\rightarrow }\widetilde{w}%
_{l}(x)=w_{kl}(x)=-w_{k}(x)+\frac{E_{l}-E_{k}}{w_{k}(x)-w_{l}(x)},
\label{transfoback2}
\end{equation}%
where $w_{kl}$ is a solution of the RS equation:

\begin{equation}
-w_{kl}^{\prime }(x)+w_{kl}^{2}(x)=\widetilde{V}_{k}(x)-E_{l}
\label{eqtransform}
\end{equation}%
with the same energy $E_{l}$ as in Eq(\ref{edr4}) but with a modified
potential $\widetilde{V}_{k}(x)=V(x)+2w_{k}^{\prime }(x)=w_{k}^{\prime
}(x)+w_{k}^{2}(x)+E_{k}$. This is the content of the finite-difference B\"{a}%
cklund algorithm\cite{Carinena,Fernandez,Mielnik,Adler1,Adler2}.

In the following we use a slightly different formulation of the B\"{a}cklund
algorithm:

If $w_{kl}(x)$ satisfy equation Eq(\ref{eqtransform}), then for every $l$
such that $E_{l}>E_{k}$:

\begin{equation}
w_{l}(x)=w_{k}(x)-\frac{E_{l}-E_{k}}{w_{k}(x)+w_{kl}(x)}.  \label{eqfond}
\end{equation}

In particular, choosing $k=0$, we have for every excited level $E_{l}>E_{0}$:

\begin{equation}
w_{l}(x)=w_{0}(x)-\frac{E_{l}-E_{0}}{w_{0}(x)+w_{0l}(x)},  \label{eqfond2}
\end{equation}%
where $w_{0l}(x)$ is a solution of:

\begin{equation}
-w_{0l}^{\prime }(x)+w_{0l}^{2}(x)=\widetilde{V}_{0}(x)-E_{l}
\label{eqtransform2}
\end{equation}%
with $\widetilde{V}_{0}(x)=V(x)+2w_{0}^{\prime }(x)$.

\section{RS functions for shape invariant potentials as terminating
continued fractions}

\subsection{Basics elements}

We recall some basic elements concerning SUSY quantum mechanics and shape
invariance\cite{Cooper,Dutt,Gendenshtein,Carinena2}.

Let $H_{-}=$ $-d^{2}/dx^{2}+V_{-}(x)$ be an hamiltonian the ground state
energy $E_{0}$ of which is supposed to be zero: $E_{0}=0$. The RS function $%
w_{0}(x)$ associated to the corresponding ground state is called the
superpotential of the system and we can write:

\begin{equation}
H_{-}=L^{+}L,
\end{equation}%
where $L=d/dx+w_{0}(x)$. The supersymmetric partner of the potential is $%
V_{-}$ defined via:

\begin{equation}
H_{+}=LL^{+}=-\frac{d^{2}}{dx^{2}}+V_{+}(x).
\end{equation}

We then have:

\begin{equation}
V_{\pm }(x)=\pm w_{0}^{\prime }(x)+w_{0}^{2}(x).  \label{susy}
\end{equation}

The potential $V_{-}(x)$ is said to be shape invariant (SIP) if it depends
upon a (multi)parameter $a\in \mathbb{R}^{N}$ and if we have the relation:

\begin{equation}
V_{+}(x,a)=V_{-}(x,f(a))+R(a)  \label{SIP}
\end{equation}%
$R\left( a\right) \in \mathbb{R}$ and $f(a)\in \mathbb{R}^{N}$ being two
given functions of $a$.

The complete spectrum of $H_{-}$ is then given by:

\begin{equation}
E_{n}(a)=R(a)+R(a_{1})+...+R(a_{n-1})=\sum_{k=0}^{n-1}R(a_{k}),
\label{spectreSIP}
\end{equation}%
where $a_{k}=f^{\left( k\right) }(a)=\overset{\text{k times}}{\overbrace{%
f\circ ...\circ f}}(a)$.

When $f$ is a simple translation $f(a)=a+\varepsilon ,\ \varepsilon \in 
\mathbb{R}^{N}$, $V_{-}$ is said to be translationally shape invariant and
we call it a TSIP.

\subsection{SIP and B\"{a}cklund algorithm}

We suppose below that the potential of the system is an SIP, $V_{-}(x,a)$,
with a discrete spectrum part $\left( E_{n}(a)\right) _{n\geq 0}$, the $%
E_{n}(a)$ forming an increasing sequence for every value of $a$ with $%
E_{0}(a)=0$.

Eq(\ref{eqtransform2}) becomes

\begin{equation}
-w_{0n}^{\prime }(x,a)+w_{0n}^{2}(x,a)=\widetilde{V}%
_{0}(x,a)-E_{n}(a)=V_{-}(x,a)+2w_{0}^{\prime }(x,a)-E_{n}(a),
\end{equation}%
that is

\begin{equation}
-w_{0n}^{\prime }(x,a)+w_{0n}^{2}(x,a)=V_{+}(x,a)-E_{n}(a).
\end{equation}

Using the shape-invariance condition Eq(\ref{SIP}), we obtain:

\begin{equation}
-w_{0n}^{\prime }(x,a)+w_{0n}^{2}(x,a)=V_{-}(x,a_{1})+R(a)-E_{n}(a),
\end{equation}%
that is, for every $m$

\begin{equation}
-\left( w_{0n}(x,a)-w_{m}(x,a_{1})\right) ^{\prime
}+w_{0n}^{2}(x,a)-w_{m}^{2}(x,a_{1})=E_{m}(a_{1})-E_{n}(a)+R(a).
\label{condSIP}
\end{equation}

If we take for $m$ the specific value $m=n-1$ and use Eq.(\ref{spectreSIP}),
Eq.(\ref{condSIP}) becomes

\begin{equation}
-\left( w_{0n}(x,a)-w_{n-1}(x,a_{1})\right) ^{\prime
}+w_{0n}^{2}(x,a)-w_{n-1}^{2}(x,a_{1})=0
\end{equation}

which is satisfied by

\begin{equation}
w_{0n}(x,a)=w_{n-1}(x,a_{1}).
\end{equation}

Eq(\ref{eqfond2}) can be now rewritten as

\begin{equation}
w_{n}(x,a)=w_{0}(x,a)-\frac{E_{n}(a)}{w_{0}(x,a)+w_{n-1}(x,a_{1})}%
=w_{0}(x,a)-\frac{\dsum\limits_{k=0}^{n-1}R(a_{k})}{%
w_{0}(x,a)+w_{n-1}(x,a_{1})}.  \label{eqfond3}
\end{equation}

By a direct iteration, using a standard notation for continued fractions, we
arrive at:

* For $n=1$:

\begin{equation}
w_{1}(x,a)=w_{0}(x,a)-\frac{R(a)}{w_{0}(x,a)+w_{0}(x,a_{1})}
\end{equation}

* For $n=2$:

\begin{equation}
w_{2}(x,a)=w_{0}(x,a)-\frac{R(a)+R(a_{1})}{w_{0}(x,a)+w_{0}(x,a_{1})-}\Rsh 
\frac{R(a_{1})}{w_{0}(x,a_{1})+w_{0}(x,a_{2})}
\end{equation}

* For $n=3$:

\begin{equation}
w_{3}(x,a)=w_{0}(x,a)-\frac{R(a)+R(a_{1})+R(a_{2})}{%
w_{0}(x,a)+w_{0}(x,a_{1})-}\Rsh \frac{R(a_{1})+R(a_{2})}{%
w_{0}(x,a_{1})+w_{0}(x,a_{2})-}\Rsh \frac{R(a_{2})}{%
w_{0}(x,a_{2})+w_{0}(x,a_{3})}
\end{equation}

...

and more generally:

\begin{equation}
w_{n}(x,a)=w_{0}(x,a)-\frac{E_{n}(a)}{w_{0}(x,a)+w_{0}(x,a_{1})-}\Rsh
...\Rsh \frac{E_{n}(a)-E_{j-1}(a)}{w_{0}(x,a_{j-1})+w_{0}(x,a_{j})-}\Rsh
...\Rsh \frac{E_{n}(a)-E_{n-1}(a)}{w_{0}(x,a_{n-1})+w_{0}(x,a_{n})}
\label{RS functions}
\end{equation}

In \cite{Kazimierz} an incomplete version of such a formula has been
outlined in a very different way. Note also that, working to connect the QHJ
formalism and SUSY quantum mechanics, Rasinariu et al.\cite{rasinariu} had
obtained the recursion relation Eq(\ref{eqfond3}) for $n=1$ in a distinct
context and again in a very different way. Their exploitation of this result
diverges from ours.

From the knowledge of the superpotential $w_{0}(x,a)$, Eq.(\ref{RS functions}%
) offers, on the basis of purely algebraic manipulations, a direct access to
the whole set of discrete excited states. It has to be compared to the known
formula\cite{Cooper,Dutt} giving the eigenstates of a SIP via the
application of differential operators (generalizing the usual
creator-annihilator ones) on the ground state:

\begin{equation}
\psi _{n}(x,a)\sim L^{+}(a)\psi _{n-1}(x,a_{1})\sim
L^{+}(a)...L^{+}(a_{n-1})\psi _{0}(x,a_{n})  \label{spectreSIP2}
\end{equation}%
with $L(a_{j})=d/dx+w_{0}(x,a_{j})$. Note that Eq(\ref{spectreSIP2}) can be
easily retrieved from the recursion relation Eq(\ref{eqfond3}). Indeed we
have from this last:

\begin{equation}
w_{n-1}(x,a_{1})=-w_{0}(x,a)+\frac{E_{n}(a)}{w_{0}(x,a)-w_{n}(x,a)}
\end{equation}%
with

\begin{equation}
E_{n}(a)=-\left( w_{0}(x,a)-w_{n}(x,a)\right) ^{\prime
}+w_{0}^{2}(x,a)-w_{n}^{2}(x,a_{1})
\end{equation}

Consequently

\begin{equation}
w_{n-1}(x,a_{1})=w_{n}(x,a)-\frac{\left( w_{0}(x,a)-w_{n}(x,a)\right)
^{\prime }}{w_{0}(x,a)-w_{n}(x,a)}
\end{equation}%
and

\begin{equation}
\psi _{n-1}(x,a_{1})\sim \left( w_{0}(x,a)-w_{n}(x,a)\right) \psi
_{n}(x,a)=L(a)\psi _{n}(x,a)
\end{equation}

In both cases the knowledge of the ground state permits a complete
reconstitution of the spectrum. Nevertheless Eq(\ref{RS functions}) avoids
to use successive differentiations and is invariant with respect to changes
of the position variable, a property which is useful in the following. We
see how the preceding results apply to simple systems, namely the
one-dimensional harmonic oscillator, the isotonic oscillator, the effective
radial potential for the Kepler problem and the Morse potential.

\subsection{One dimensional harmonic oscillator}

The one dimensional harmonic oscillator potential is\cite{Cooper}:

\begin{equation}
V_{-}(x)=\frac{\omega ^{2}}{4}x^{2}+V_{0}
\end{equation}%
and the RS equation for the ground state $\left( E_{0}=0\right) $ is

\begin{equation}
-w_{0}^{\prime }(x)+w_{0}^{2}(x)=\frac{\omega ^{2}}{4}x^{2}+V_{0}
\label{potOH}
\end{equation}

or

\begin{equation}
\left( \frac{d}{dx}-\left( w_{0}(x)+\frac{\omega }{2}x\right) \right) \left(
w_{0}(x)-\frac{\omega }{2}x\right) =-V_{0}-\frac{\omega }{2}
\end{equation}

Choosing $V_{0}=-\frac{\omega }{2}$, this equation admits clearly a
polynomial solution:

\begin{equation}
w_{0}(x)=\frac{\omega }{2}x  \label{fondOH}
\end{equation}%
which is such that $w^{\prime }(x)>0$ between the turning points of the
classical motion (for a bound state we must have in this domain $E<V(x)$).
As required, the ground-state eigenfunction associated to this
superpotential does not present any nodes on the real line.

The SUSY partner of $V_{-}(x)$ is then:

\begin{equation}
V_{+}(x)=w_{0}^{\prime }(x)+w_{0}^{2}(x)=\frac{\omega ^{2}}{4}x^{2}+\frac{%
\omega }{2}=V_{-}(x)+\omega
\end{equation}

We recognize a SIP (cf Eq(\ref{SIP})) characterized by:

\begin{equation}
a=\frac{\omega }{2}=\ f(a),\ a_{k}=a,\ R(a)=2a=\omega  \label{SIPOH}
\end{equation}%
that is a TSIP with a zero translation amplitude.

Consequently:

\begin{equation}
\left\{ 
\begin{array}{c}
E_{n}(a)-E_{j-1}(a)=\dsum\limits_{k=j-1}^{n-1}\omega =\left( n-j+1\right)
\omega \\ 
w_{0}(x,a_{j-1})+w_{0}(x,a_{j})=\omega x%
\end{array}%
\right.
\end{equation}%
and Eq(\ref{RS functions}) becomes in this case:

\begin{equation}
w_{n}(x,\omega )=\frac{\omega }{2}x-\frac{n\omega }{\omega x-}\Rsh ...\Rsh 
\frac{\left( n-j+1\right) \omega }{\omega x-}\Rsh ...\Rsh \frac{1}{x}.
\label{RS functions OH}
\end{equation}

In particular:

\begin{equation}
\left\{ 
\begin{array}{c}
w_{1}(x)=\frac{\omega }{2}x-\frac{1}{x} \\ 
\\ 
w_{2}(x)=\frac{\omega }{2}x-\frac{2\omega }{\omega x-1/x}=\frac{\omega }{2}x-%
\frac{2\omega x}{\omega x^{2}-1} \\ 
\\ 
w_{3}(x)=\frac{\omega }{2}x-\frac{3\omega }{\omega x-2\omega /\left( \omega
x-1/x\right) }=\frac{\omega }{2}x-\frac{3\left( \omega x^{2}-1\right) }{%
x\left( \omega x^{2}-3\right) }%
\end{array}%
\right.  \label{soledrOH2}
\end{equation}%
from which, with Eq(\ref{fondOH}), we deduce immediately the corresponding
four first eigenstates:

\begin{equation}
\left\{ 
\begin{array}{c}
\psi _{0}(x)=\exp \left( -\int^{x}w_{0}(s)ds\right) \sim \exp \left( -\frac{%
\omega }{4}x^{2}\right) \\ 
\\ 
\psi _{1}(x)=\exp \left( -\int^{x}w_{1}(s)ds\right) \sim \exp \left( -\frac{%
\omega }{4}x^{2}\right) \exp \left( \int^{x}\frac{1}{s}ds\right) =x\exp
\left( -\frac{\omega }{4}x^{2}\right) \\ 
\\ 
\psi _{2}(x)=\exp \left( -\int^{x}w_{2}(s)ds\right) \sim \exp \left( -\frac{%
\omega }{4}x^{2}\right) \exp \left( \int^{x}\frac{2\omega s}{\omega s^{2}-1}%
ds\right) =\left( \omega x^{2}-1\right) \exp \left( -\frac{\omega }{4}%
x^{2}\right) \\ 
\\ 
\psi _{3}(x)=\exp \left( -\int^{x}w_{3}(s)ds\right) \sim \exp \left( -\frac{%
\omega }{4}x^{2}\right) \exp \left( \int^{x}\frac{3\left( \omega
s^{2}-1\right) }{\omega s^{3}-3s}ds\right) =\left( \omega x^{3}-3x\right)
\exp \left( -\frac{\omega }{4}x^{2}\right) .%
\end{array}%
\right.  \label{foOH}
\end{equation}

\subsection{Isotonic oscillator}

The isotonic oscillator potential is\cite{weissman,zhu}:

\begin{equation}
V_{-}(x)=\frac{\omega ^{2}}{4}x^{2}+\frac{l(l+1)}{x^{2}}+V_{0},\ l>0
\end{equation}

and the RS equation for the ground state $\left( E_{0}=0\right) $:

\begin{equation}
-w_{0}^{\prime }(x)+w_{0}^{2}(x)=\frac{\omega ^{2}}{4}x^{2}+\frac{l(l+1)}{%
x^{2}}+V_{0}  \label{potIsot}
\end{equation}

Looking for a solution of the form $w_{0}(x)=\lambda x-\mu /x$, we obtain
immediately that Eq(\ref{potIsot}) is satisfied when $\lambda =\omega /2,\
\mu =l+1$ and $V_{0}=-\omega \left( l+\frac{3}{2}\right) $, giving for the
superpotential:

\begin{equation}
w_{0}(x)=\frac{\omega }{2}x-\frac{l+1}{x}  \label{fondIsot}
\end{equation}

The signs have been chosen in order that $w^{\prime }(x)>0$ between the
turning points of the classical motion.

The SUSY partner of $V_{-}(x)$ is then:

\begin{equation}
V_{+}(x)=w_{0}^{\prime }(x)+w_{0}^{2}(x)=\frac{\omega ^{2}}{4}x^{2}+\frac{%
(l+2)(l+1)}{x^{2}}\frac{\omega }{2}-\omega \left( l+\frac{3}{2}\right)
+\omega .
\end{equation}

We recognize a TSIP (cf Eq(\ref{SIP})) characterized by:

\begin{equation}
a=\left( \frac{\omega }{2},l+1\right) ,\ f(a)=\left( \frac{\omega }{2}%
,l+2\right) ,\ a_{k}=\left( \frac{\omega }{2},l+1+k\right) ,\ R(a)=2\omega .
\label{SIPIsot}
\end{equation}

Consequently

\begin{equation}
\left\{ 
\begin{array}{c}
E_{n}(a)-E_{j-1}(a)=\dsum\limits_{k=j-1}^{n-1}2\omega =2\left( n-j+1\right)
\omega \\ 
w_{0}(x,a_{j-1})+w_{0}(x,a_{j})=\omega x-\left( 2\left( l+j\right) +1\right)
/x%
\end{array}%
\right.
\end{equation}%
and Eq(\ref{RS functions}) becomes in this case:

\begin{equation}
w_{n}(x,a)=\frac{\omega }{2}x-\frac{l+1}{x}-\frac{2n\omega }{\omega x-\left(
2l+3\right) /x-}\Rsh ...\Rsh \frac{2\left( n-j+1\right) \omega }{\omega
x-\left( 2\left( l+j\right) +1\right) /x-}\Rsh ...\Rsh \frac{2\omega }{%
\omega x-\left( 2\left( l+n\right) -1\right) /x}  \label{RS functions Isot}
\end{equation}

with in particular:

\begin{equation}
w_{1}(x,a)=\frac{\omega }{2}x-\frac{l+1}{x}-\frac{2\omega }{\omega x-\left(
2l+3\right) /x}.  \label{soledrIsot2}
\end{equation}%
From Eq(\ref{fondIsot}) and Eq(\ref{soledrIsot2}) we deduce immediately the
corresponding two first eigenstates:

\begin{equation}
\left\{ 
\begin{array}{c}
\psi _{0}(x)=\exp \left( -\int^{x}w_{0}(s,a)ds\right) \sim \exp \left( -%
\frac{\omega }{4}x^{2}\right) \exp \left( \left( l+1\right) \int^{x}\frac{1}{%
s}ds\right) =x^{l+1}\exp \left( -\frac{\omega }{4}x^{2}\right) \\ 
\\ 
\psi _{1}(x)=\exp \left( -\int^{x}w_{1}(s,a)ds\right) \sim x^{l+1}\exp
\left( -\frac{\omega }{4}x^{2}\right) \exp \left( \int^{x}\frac{2\omega }{%
\omega s-\left( 2l+3\right) /s}ds\right) =x^{l+1}\left( \omega
x^{2}-(2l+3)\right) \exp \left( -\frac{\omega }{4}x^{2}\right) .%
\end{array}%
\right.  \label{foIsot}
\end{equation}

\subsection{Effective radial potential for the Kepler-Coulomb problem}

The effective radial potential for the Kepler-Coulomb is\cite{Cooper}:

\begin{equation}
V_{-}(x)=-\frac{\gamma }{x}+\frac{l(l+1)}{x^{2}}+V_{0},\ k>0  \label{potkep}
\end{equation}

Defining $y=1/x$, $V_{-}$ takes the form:

\begin{equation}
V_{-}(y)=-\gamma y+l(l+1)y^{2}+V_{0},\ k>0
\end{equation}

and the RS equation for the ground state $\left( E_{0}=0\right) $:

\begin{equation}
y^{2}w_{0}^{\prime }(y)+w_{0}^{2}(y)=-\gamma y+l(l+1)y^{2}+V_{0}
\end{equation}

Looking for a solution of the form $w_{0}(x)=-\lambda y+\mu $, we obtain
immediately that Eq(\ref{potIsot}) is satisfied when $\lambda =l+1,\ \mu
=\gamma /2(l+1)$ and $V_{0}=\gamma ^{2}/4(l+1)^{2}$, giving for the
superpotential:

\begin{equation}
w_{0}(y)=-\left( l+1\right) y+\frac{\gamma }{2(l+1)}  \label{fondkep}
\end{equation}

The signs have been chosen in order that $w^{\prime }(x)>0$ between the
turning points of the classical motion.

The SUSY partner of $V_{-}(x)$ is then:

\begin{equation}
V_{+}(y)=-2y^{2}w_{0}^{\prime }(y)+V_{-}(y)=-\gamma y+(l+1)\left( l+2\right)
y^{2}+\frac{\gamma ^{2}}{4(l+1)^{2}}
\end{equation}

We recognize a TSIP (cf Eq(\ref{SIP})) characterized by:

\begin{equation}
a=l+1,\ f(a)=l+2,\ a_{k}=l+1+k,\ R(a_{k})=\frac{\gamma ^{2}}{4a_{k}^{2}}-%
\frac{\gamma ^{2}}{4a_{k+1}^{2}}.
\end{equation}

Consequently

\begin{equation}
E_{n}(a)=\sum_{k=0}^{n-1}R(a_{k})=\frac{\gamma ^{2}}{4a^{2}}-\frac{\gamma
^{2}}{4a_{n}^{2}}=\frac{\gamma ^{2}}{4\left( l+1\right) ^{2}}-\frac{\gamma
^{2}}{4\left( l+1+n\right) ^{2}}
\end{equation}%
and

\begin{equation}
\left\{ 
\begin{array}{c}
E_{n}(a)-E_{j-1}(a)=\frac{\gamma ^{2}}{4a_{j-1}^{2}}-\frac{\gamma ^{2}}{%
4a_{n}^{2}}=\frac{\gamma ^{2}}{4}\left( \frac{1}{\left( l+j\right) ^{2}}-%
\frac{1}{\left( l+1+n\right) ^{2}}\right) \\ 
\\ 
w_{0}(x,a_{j-1})+w_{0}(x,a_{j})=-\left( a_{j-1}+a_{j}\right) y+\frac{\gamma 
}{2}\left( \frac{1}{a_{j-1}}+\frac{1}{a_{j}}\right) =-2\left( l+j+\frac{1}{2}%
\right) y+\frac{\gamma }{2}\left( \frac{1}{l+j}+\frac{1}{l+1+j}\right)%
\end{array}%
\right.
\end{equation}

Eq(\ref{RS functions}) becomes in this case:

\begin{eqnarray}
w_{n}(y,a) &=&-ay+\frac{\gamma }{2a}-\frac{\gamma ^{2}/4a^{2}-\gamma
^{2}/4a_{n}^{2}}{-\left( a+a_{1}\right) y+\gamma /2\left( 1/a+1/a_{1}\right)
-}\Rsh \\
... &\Rsh &\frac{\gamma ^{2}/4a_{j-1}^{2}-\gamma ^{2}/4a_{n}^{2}}{-\left(
a_{j-1}+a_{j}\right) y+\gamma /2\left( 1/a_{j-1}+1/a_{j}\right) -}\Rsh 
\notag \\
... &\Rsh &\frac{\gamma ^{2}/4a_{n-1}^{2}-\gamma ^{2}/4a_{n}^{2}}{-\left(
a_{n-1}+a_{n}\right) y+\gamma /2\left( 1/a_{n-1}+1/a_{n}\right) }  \notag
\end{eqnarray}

In particular:

\begin{equation}
w_{1}(y,a)=w_{0}(y)+\left( \frac{\gamma }{2aa_{1}}\right) ^{2}\frac{1}{%
y-\gamma /2aa_{1}}  \label{soledrkep}
\end{equation}%
or

\begin{equation}
w_{1}(x,a)=w_{0}(x)+\frac{1}{2aa_{1}/\gamma -x}-\frac{\gamma }{2aa_{1}}
\label{soledrkep2}
\end{equation}

From Eq(\ref{fondkep}) and Eq(\ref{soledrkep2}) we deduce immediately the
corresponding two first eigenstates:

\begin{eqnarray}
\psi _{0}(x) &=&\exp \left( -\int w_{0}(x,a)dx\right) =\exp \left( \int
\left( \frac{\left( l+1\right) }{x}-\frac{\gamma }{2(l+1)}\right) dx\right)
\\
&\sim &x^{l+1}\exp \left( -\frac{\gamma }{2(l+1)}x\right)
\end{eqnarray}%
and:

\begin{eqnarray}
\psi _{1}(x) &=&\exp \left( -\int w_{1}(x,a)dx\right) =\exp \left( -\int
w_{0}(x,a)dx\right) \exp \left( \int \left( \frac{1}{x-2aa_{1}/\gamma }+%
\frac{\gamma }{2aa_{1}}\right) dx\right) \\
&\sim &x^{l+1}\left( x-\frac{2\left( l+1\right) \left( l+2\right) }{\gamma }%
\right) \exp \left( -\frac{\gamma }{2(l+2)}x\right) .  \notag
\end{eqnarray}

\subsection{Morse potential}

The Morse potential is\cite{Cooper}:

\begin{equation}
V_{-}(x)=A^{2}+B^{2}e^{-2\alpha x}-2B\left( A+\frac{\alpha }{2}\right)
e^{-\alpha x},\ \alpha >0  \label{potMorse}
\end{equation}

Defining $y=\exp \left( -\alpha x\right) $, $V_{-}$ takes the form:

\begin{equation}
V_{-}(y)=\left( By-A\right) ^{2}-\alpha By.
\end{equation}

In terms of the $y$ variable, the associated RS equation is:

\begin{equation*}
-\alpha yw_{n}^{\prime }(y)=E_{n}-\left( By-A\right) ^{2}+\alpha
By+w_{n}^{2}(y)
\end{equation*}%
and for the ground state, $E_{0}=0$, we have

\begin{equation}
\left( \alpha y\frac{d}{dy}+\left( w_{0}(y)-\left( By-A\right) \right)
\right) \left( w_{0}(y)+\left( By-A\right) \right) =0
\end{equation}%
a solution of which is immediately obtained as

\begin{equation}
w_{0}(y)=-By+A  \label{fondmorse}
\end{equation}

The SUSY partner of $V_{-}(x)$ is then:

\begin{equation}
V_{+}(y)=-\alpha yw_{0}^{\prime }(y)+w_{0}^{2}(y)=\left( By-A+\alpha \right)
^{2}-\alpha By+\alpha ^{2}-2\alpha A
\end{equation}

We recognize a TSIP (cf Eq(\ref{SIP})) characterized by:

\begin{equation}
a=A,\ f(a)=A-\alpha ,\ R(a)=\alpha ^{2}-2\alpha =a^{2}-a_{1}^{2}
\label{SIPMorse}
\end{equation}

Consequently

\begin{equation}
E_{n}(a)=\sum_{k=0}^{n-1}R(a_{k})=a^{2}-a_{n}^{2}=n\alpha \left( 2a-n\alpha
\right)
\end{equation}%
and

\begin{equation}
\left\{ 
\begin{array}{c}
E_{n}(a)-E_{j-1}(a)=\dsum\limits_{k=j-1}^{n-1}\left(
a_{k}^{2}-a_{k+1}^{2}\right) =a_{j-1}^{2}-a_{n}^{2} \\ 
w_{0}(x,a_{j-1})+w_{0}(x,a_{j})=a_{j-1}+a_{j}-2By%
\end{array}%
\right.
\end{equation}

Eq(\ref{RS functions}) becomes in this case:

\begin{equation}
w_{n}(y,a)=a-By-\frac{a^{2}-a_{n}^{2}}{a+a_{1}-2By-}\Rsh ...\Rsh \frac{%
a_{j-1}^{2}-a_{n}^{2}}{a_{j-1}+a_{j}-2By-}\Rsh ...\Rsh \frac{%
a_{n-1}^{2}-a_{n}^{2}}{a_{n-1}+a_{n}-2By}  \label{RS functions Morse}
\end{equation}

In particular:

\begin{equation}
\left\{ 
\begin{array}{c}
w_{1}(y,a)=a-By-\frac{a^{2}-a_{1}^{2}}{a+a_{1}-2By} \\ 
\\ 
w_{2}(y,a)=a-By-\frac{a^{2}-a_{2}^{2}}{a+a_{1}-2By-\left(
a_{1}^{2}-a_{2}^{2}\right) /\left( a_{1}+a_{2}-2By\right) }%
\end{array}%
\right.  \label{soledrMorse2}
\end{equation}

From Eq(\ref{fondmorse}) and Eq(\ref{soledrMorse2}) we deduce immediately
the corresponding two first eigenstates:

\begin{equation}
\psi _{0}(x)=\exp \left( -\int w_{0}(x,a)dx\right) =\exp \left( \frac{1}{%
\alpha }\int \frac{w_{0}(y,a)}{y}dy\right) \sim y^{A/\alpha }\exp \left( -%
\frac{B}{\alpha }y\right) =\exp \left( -Ax-\frac{B}{\alpha }\exp \left(
-\alpha x\right) \right)  \label{foMorse}
\end{equation}%
and:

\begin{eqnarray}
\psi _{1}(x) &=&\exp \left( -\int w_{1}(x,a)dx\right) =\exp \left( \frac{1}{%
\alpha }\int \frac{w_{1}(y,a)}{y}dy\right) \sim y^{\left( A-\alpha \right)
/\alpha }\exp \left( -\frac{B}{\alpha }y\right) \left( By-\left( A-\frac{1}{2%
}\alpha \right) \right)  \label{foMorse2} \\
&\sim &\exp \left( -\left( A-\alpha \right) x-\frac{B}{\alpha }e^{-\alpha
x}\right) \left( Be^{-\alpha x}-\left( A-\frac{1}{2}\alpha \right) \right) .
\notag
\end{eqnarray}

\section{Potentials of first and second categories}

\subsection{Definitions}

\subsubsection{First category}

We say that a one-dimensional potential is of first category if there exists
a change of variable $x\rightarrow u$ transforming the potential into an
harmonic one $V(x)\rightarrow V(u)=\widetilde{\lambda }_{2}u^{2}+\widetilde{%
\lambda }_{1}u+\widetilde{\lambda }_{0}$, such that $u(x)$ satisfies a
constant coefficient Riccati equation:

\begin{equation}
\frac{du(x)}{dx}=A_{0}+A_{1}u(x)+A_{2}u^{2}(x),  \label{potreg}
\end{equation}%
$du(x)/dx$ being of constant sign in all the range of values of $x$ and $u$.

The one-dimensional harmonic oscillator corresponds to the special case $%
A_{1}=A_{2}=0$ and the Morse potential is generically associated to the case 
$A_{1}\neq 0,\ A_{2}=0$. These two examples have been treated above.

We note $\Delta =A_{1}^{2}-4A_{2}A_{0}$ the discriminant of the right hand
side polynomial in Eq(\ref{potreg}).

If $A_{2}\neq 0,\ \Delta \neq 0$, it is always possible, via an affine
change of variable $u=ay+b$, to\ reduce Eq(\ref{potreg}) to the canonical
form:

\begin{equation}
\frac{dy}{dx}=\alpha \pm \alpha y^{2}(x)>0.
\end{equation}

If $A_{2}\neq 0,\ \Delta =0$, a straightforward affine change of variable
brings Eq(\ref{potreg}) to the form:

\begin{equation}
\frac{dy}{dx}=\alpha y^{2}(x).
\end{equation}%
that is $y=\frac{1}{\alpha \left( x-x_{0}\right) }$ which corresponds to the
radial effective Kepler-Coulomb potential studied above.

In the following we then consider that all the first category potentials are
of the type:

\begin{equation}
\left\{ 
\begin{array}{c}
V(x)\rightarrow V(y)=\lambda _{2}y^{2}+\lambda _{1}y+\lambda _{0},\ \lambda
_{2}>0 \\ 
dy/dx=\alpha \pm \alpha y^{2}(x)>0.%
\end{array}%
\right.  \label{1st category}
\end{equation}

The new variable is then $y=\tan \left( \alpha x+\varphi _{0}\right) $ for
the positive sign and $y=\tanh \left( \alpha x+\varphi _{0}\right) $ or $%
y=\coth \left( \alpha x+\varphi _{0}\right) $ (depending on the sign of $%
\alpha $) for the negative sign.

In this category we find all the potentials listed in the Annexe A which
contains also the three exceptional cases corresponding to the harmonic,
Morse and effective radial Kepler-Coulomb potentials.

\subsubsection{Second category}

We say that a one dimensional potential is of second category if there
exists a change of variable $x\rightarrow u$ transforming the potential into
an isotonic one $V(x)\rightarrow V(u)=\widetilde{\lambda }_{2}u^{2}+%
\widetilde{\lambda }_{0}+\widetilde{\mu }_{2}/u^{2}$, such that $u(x)$
satisfies a constant coefficient Riccati equation of the form

\begin{equation}
\frac{du(x)}{dx}=A_{0}+A_{2}u^{2}(x)  \label{potsing}
\end{equation}%
$du(x)/dx$ being of constant sign in all the range of values of $x$ and $u$.

When $A_{2}$ and $A_{0}$ are nonzero, a linear change of variable $u=ay$ \
reduces Eq(\ref{potsing}) to the canonical form:

\begin{equation}
\left\{ 
\begin{array}{c}
V(x)\rightarrow V(y)=\lambda _{2}y^{2}+\frac{\mu _{2}}{y^{2}}+\lambda _{0},\
\lambda _{2},\mu _{2}>0 \\ 
dy/dx=\alpha \pm \alpha y^{2}(x)>0.%
\end{array}%
\right. ,  \label{2 nd category}
\end{equation}

Again, the new variable is then $y=\tan \left( \alpha x+\varphi _{0}\right) $
for the positive sign and $y=\tanh \left( \alpha x+\varphi _{0}\right) $ for
the negative sign (due to the $y\rightarrow 1/y$ symmetry of the functional
form of $V$, we can limit ourselves to this two solutions).

The example of the isotonic potential, treated above, is generically
associated to the cases $A_{0}=0$ or $A_{2}=0$.

In this category we find all the potentials listed in the Annexe B.

\section{ Polynomial solutions for the RS equations associated to first and
second category potentials}

\subsection{First category}

The RS equation associated to a first category potential $V(x)$ is in terms
of the $y$ variable(see Eq.(\ref{1st category})):

\begin{equation}
-\left( \alpha _{0}+\alpha _{1}y+\alpha _{2}y^{2}\right) w^{\prime
}(y)+w^{2}(y)=\lambda _{2}y^{2}+\lambda _{1}y+\lambda _{0}-E
\label{edrTSIP1}
\end{equation}

($\alpha _{0}=\alpha =\pm \alpha _{2}$ , $\alpha _{1}=0$).

We look for a polynomial solution of degree $N$ of this equation, describing
a bound state of the system. Inserting

\begin{equation}
w(y)=\sum_{n=0}^{N}b_{n}y^{n},\ b_{N}\neq 0
\end{equation}%
into Eq.(\ref{edrTSIP1}), we obtain:

\begin{equation}
-\sum_{l=0}^{N+1}\left( \sum_{\substack{ n=0  \\ n+m=l}}^{N-1}\sum_{m=0}^{2}%
\left( n+1\right) \alpha _{m}b_{n+1}\right) y^{l}+\sum_{l=0}^{2N}\left( \sum 
_{\substack{ n,m=0  \\ n+m=l}}^{N}b_{n}b_{m}\right)
y^{l}=\sum_{n=0}^{2}\left( \lambda _{n}-E\delta _{n,0}\right) y^{n}.
\label{eqpoly1}
\end{equation}

Since $b_{N}\neq 0$, we must have $N+1=2N\Leftrightarrow N=1$ or $%
2=2N\Leftrightarrow N=1$. The polynomial is necessarily of degree 1:

\begin{equation}
w(y)=b_{0}+b_{1}y.  \label{solpolyedr}
\end{equation}

Since $y(x)$ satisfies a constant coefficient Riccati equation, we recover
the first Gendenshtein ansatz\cite{Gendenshtein}.

For a bound state $E<V(x)$, that is $w^{\prime }(x)>0$, between the turning
points of the classical motion and $b_{1}$ is of the same sign as $y^{\prime
}(x)$, that is positive. In this domain $w(x)$ is not singular and the wave
function does not present any node, which corresponds to the ground state.

Inserting Eq.(\ref{solpolyedr}) into Eq.(\ref{eqpoly1}) we obtain for the
unknown coefficients $b_{0}$, $b_{1}$ a set of 3 equations:

\begin{equation}
\left\{ 
\begin{array}{c}
-\alpha b_{1}+b_{0}^{2}=\lambda _{0}-E \\ 
\\ 
2b_{0}b_{1}=\lambda _{1} \\ 
\\ 
\mp \alpha b_{1}+b_{1}^{2}=\lambda _{2}>0%
\end{array}%
\right.
\end{equation}%
which imply a constraint on $E$ fixing the value of the ground state energy.

After straightforward calculations we arrive at

\begin{equation}
\left\{ 
\begin{array}{c}
b_{0}=\lambda _{1}/2\beta _{\pm }\left( \lambda _{2}\right) \\ 
b_{1}=\beta _{\pm }\left( \lambda _{2}\right) >0%
\end{array}%
\right.
\end{equation}%
with

\begin{equation}
\beta _{\pm }\left( \lambda \right) =\pm \alpha /2+\sqrt{\left( \alpha
/2\right) ^{2}+\lambda }.  \label{beta+}
\end{equation}%
The ground state energy is given by

\begin{equation}
E_{0}=\lambda _{0}\pm \alpha \beta _{\pm }\left( \lambda _{2}\right) -\left(
\lambda _{1}/2\beta _{\pm }\left( \lambda _{2}\right) \right) ^{2}
\label{fond1}
\end{equation}%
and the corresponding RS function:

\begin{equation}
w_{0}(y)=\frac{\lambda _{1}}{2\beta _{\pm }\left( \lambda _{2}\right) }%
+\beta _{\pm }\left( \lambda _{2}\right) y  \label{superpot1}
\end{equation}

\subsection{Second category}

The RS equation associated to a first category potential $V(x)$ is in terms
of the $y$ variable(see Eq.(\ref{2 nd category})):

\begin{equation}
-\left( \alpha _{0}+\alpha _{1}y+\alpha _{2}y^{2}\right) w^{\prime
}(y)+w^{2}(y)=\lambda _{2}y^{2}+\lambda _{0}+\frac{\mu _{2}}{y^{2}}-E
\label{edrTSIP2}
\end{equation}%
($\alpha _{0}=\alpha =\pm \alpha _{2}$ , $\alpha _{1}=0$).

We look for a solution of Eq.(\ref{edrTSIP2}) which is a Laurent polynomial,
that is a polynomial in $y$ and $1/y$. Inserting

\begin{equation}
w(y)=\sum_{n=-M}^{N}b_{n}y^{n},\ b_{N},b_{-M}\neq 0
\end{equation}%
into Eq.(\ref{edrTSIP2}) we obtain:

\begin{equation}
-\sum_{l=-M+1}^{N+1}\left( \sum_{\substack{ n=-M-1  \\ n+m=l}}%
^{N-1}\sum_{m=0}^{2}\left( n+1\right) A_{m}b_{n+1}\right)
y^{l}+\sum_{l=-2M}^{2N}\left( \sum_{\substack{ n,m=-M  \\ n+m=l}}%
^{N}b_{n}b_{m}\right) y^{l}=\lambda _{2}y^{2}+\lambda _{0}+\frac{\mu _{2}}{%
y^{2}}-E  \label{eqpoly2}
\end{equation}

Since $b_{N},b_{-M}\neq 0$, we must have $N+1=2N\Leftrightarrow N=1$ or $%
2=2N\Leftrightarrow N=1$ and $2M=M-1\Leftrightarrow M=1$ or $%
2=2M\Leftrightarrow M=1$. The Laurent polynomial is then of degree 1, as for
the regular part as for the singular part:

\begin{equation}
w(y)=b_{0}+b_{1}y+\frac{b_{-1}}{y}.  \label{solpolyedr2}
\end{equation}

Since $y(x)$ satisfies now a constant coefficients Riccati equation without
a term of first degree, we recover the second Gendenshtein ansatz\cite%
{Gendenshtein}.

For a bound state $E<V(x)$, that is $w^{\prime }(x)>0$, between the turning
points of the classical motion and $b_{1}-b_{-1}/y^{2}$ is of the same sign
as $y^{\prime }(x)$, that is positive. This necessitates that $b_{1}$ is
positive and $b_{-1}$ negative. In this domain $w(x)$ is not singular and
the wave function does not present any node, which corresponds to the ground
state.

Inserting Eq.(\ref{solpolyedr2}) into Eq.(\ref{eqpoly2}), we obtain for the
unknown coefficients $b_{0}$, $b_{1}$ and $b_{-1}$ a set of 4 equations:

\begin{equation}
\left\{ 
\begin{array}{c}
E-\lambda _{0}=\alpha b_{1}-b_{0}^{2}\mp \alpha b_{-1}-2b_{1}b_{-1} \\ 
\\ 
b_{0}b_{-1}=b_{0}b_{1}=0 \\ 
\\ 
b_{1}^{2}\mp b_{1}\alpha =\lambda _{2}>0 \\ 
\\ 
b_{-1}^{2}+b_{-1}\alpha =\mu _{2}>0%
\end{array}%
\right.  \label{syst 2}
\end{equation}%
which imply a constraint on $E$, fixing the value of the ground state
energy.\bigskip

Eq.(\ref{syst 2}) gives (see Eq.(\ref{beta+})):

\begin{equation}
\left\{ 
\begin{array}{c}
b_{0}=0 \\ 
b_{1}=\beta _{\pm }\left( \lambda _{2}\right) >0 \\ 
b_{-1}=-\beta _{+}\left( \mu _{2}\right) <0.%
\end{array}%
\right.
\end{equation}

The ground-state energy is given by:

\begin{equation}
E_{0}=\lambda _{0}+\alpha \left( \beta _{\pm }\left( \lambda _{2}\right) \pm
\beta _{+}\left( \mu _{2}\right) \right) +2\beta _{\pm }\left( \lambda
_{2}\right) \beta _{+}\left( \mu _{2}\right)  \label{fond2}
\end{equation}%
and the corresponding RS function is

\begin{equation}
w_{0}(y)=\beta _{\pm }\left( \lambda _{2}\right) y-\frac{\beta _{+}\left(
\mu _{2}\right) }{y}  \label{superpot2}
\end{equation}

\section{\protect\bigskip\ Shape invariance properties of first and second
category potentials}

\subsection{First category}

We consider a first category potential (see Eq.(\ref{1st category})) $%
V_{-}(y)=\lambda _{2}y^{2}+\lambda _{1}y+\lambda _{0},\ \lambda _{2}>0$, for
which $dy/dx=\alpha \pm \alpha y^{2}(x)>0$ and $E_{0}=0$. This last
constraint implies (see Eq.(\ref{fond1})) that

\begin{equation*}
\lambda _{0}=\lambda _{0}\left( \lambda _{2}\right) =\left( \frac{\lambda
_{1}}{2\beta _{\pm }\left( \lambda _{2}\right) }\right) ^{2}-\alpha \beta
_{\pm }\left( \lambda _{2}\right) .
\end{equation*}

The SUSY partner of $V_{-}(y)$ is

\begin{eqnarray}
V_{+}(y) &=&V_{-}(y)+2w_{0}^{\prime }(x)=V_{-}(y)+2w_{0}^{\prime }(y)\frac{dy%
}{dx} \\
&=&y^{2}\left( \lambda _{2}\pm 2\alpha \beta _{\pm }\left( \lambda
_{2}\right) \right) +\lambda _{1}y+\lambda _{0}\left( \lambda _{2}\right)
+2\alpha \beta _{\pm }\left( \lambda _{2}\right) .
\end{eqnarray}

We make the following change of parameter:

\begin{equation}
a=\beta _{\pm }\left( \lambda _{2}\right)
\end{equation}%
in which case

\begin{equation*}
\lambda _{2}=a\left( a\mp \alpha \right) ,\ \lambda _{0}(a)=\left( \frac{%
\lambda _{1}}{2a}\right) ^{2}-\alpha a.
\end{equation*}%
The initial potential is now

\begin{equation}
V_{-}(y,a)=a\left( a\mp \alpha \right) y^{2}+\lambda _{1}y+\lambda _{0}(a)
\end{equation}%
and its SUSY partner is

\begin{eqnarray}
V_{+}(y,a) &=&y^{2}a\left( a\pm \alpha \right) +\lambda _{1}y+\lambda
_{0}\left( a\right) +2\alpha a \\
&=&y^{2}a_{1}\left( a_{1}\mp \alpha \right) +\lambda _{1}y+\lambda
_{0}\left( a_{1}\right) +R(a)  \notag
\end{eqnarray}%
with:

\begin{equation}
\ \left\{ 
\begin{array}{c}
a_{1}=a\pm \alpha \\ 
\ R(a)=\lambda _{0}\left( a\right) -\lambda _{0}\left( a_{1}\right) +2\alpha
a=\phi _{1,\pm }\left( a\right) -\phi _{1,\pm }\left( a_{1}\right) ,%
\end{array}%
\right.
\end{equation}%
where

\begin{equation}
\phi _{1,\pm }\left( a\right) =\mp a^{2}+\frac{\lambda _{1}^{2}}{4a^{2}}
\end{equation}

We recognize (see Eq.(\ref{SIP})) the characteristic behaviour of a TSIP.
More generally

\begin{equation}
\left\{ 
\begin{array}{c}
a_{k}=a\pm k\alpha \\ 
R(a_{k})=\phi _{1,\pm }\left( a_{k}\right) -\phi _{1,\pm }\left(
a_{k+1}\right) ,%
\end{array}%
\right.
\end{equation}

The spectrum (see Eq.(\ref{spectreSIP})) becomes%
\begin{equation}
E_{n}(a)=\phi _{1,\pm }\left( a\right) -\phi _{1,\pm }\left( a_{n}\right)
\label{spectrecat 1}
\end{equation}%
or

\begin{equation}
E_{n}(a)=\phi _{1,\pm }\left( a\right) \pm \alpha ^{2}\left( n\pm \frac{a}{%
\alpha }\right) ^{2}-\frac{\lambda _{1}^{2}}{4\alpha ^{2}}\frac{1}{\left(
n\pm a/\alpha \right) ^{2}}  \label{spectrecat1}
\end{equation}

$E_{n}$ is then an isotonic function of $n\pm a/\alpha $. As for the
corresponding superpotential, it takes the form (see Eq.(\ref{superpot1})):

\begin{equation}
w_{0}(y,a)=ay+\frac{\lambda _{1}}{2a},\ a>0  \label{superpotcat1}
\end{equation}

Using Eq.(\ref{RS functions}), Eq.(\ref{spectrecat 1}) and Eq.(\ref%
{superpotcat1}) we deduce finally the general form for the RS function
associated to the n$^{th}$ level of the spectrum of an arbitrary potential
of the first category:

\begin{eqnarray}
w_{n}(y,a) &=&ay+\frac{\lambda _{1}}{2a}-\frac{\phi _{1,\pm }\left( a\right)
-\phi _{1,\pm }\left( a_{n}\right) }{\left( a+a_{1}\right) y+\frac{\lambda
_{1}}{2}\left( 1/a+1/a_{1}\right) -}\Rsh ...  \label{RS functions cat1} \\
&&  \notag \\
&\Rsh &\frac{\phi _{1,\pm }\left( a_{j-1}\right) -\phi _{1,\pm }\left(
a_{n}\right) }{\left( a_{j-1}+a_{j}\right) y+\frac{\lambda _{1}}{2}\left(
1/a_{j-1}+1/a_{j}\right) -}\Rsh ...  \notag \\
&&  \notag \\
&\Rsh &\frac{\phi _{1,\pm }\left( a_{n-1}\right) -\phi _{1,\pm }\left(
a_{n}\right) }{\left( a_{n}+a_{n-1}\right) y+\frac{\lambda _{1}}{2}\left(
1/a_{n-1}+1/a_{n}\right) }  \notag
\end{eqnarray}

\subsection{Second category}

We consider a second category potential (see Eq.(\ref{2 nd category})) $%
V_{-}(y)=\lambda _{2}y^{2}+\lambda _{0}+\frac{\mu _{2}}{y^{2}}$, for which $%
dy/dx=\alpha \pm \alpha y^{2}(x)>0$ and $E_{0}=0$. This last constraint
implies (see Eq.(\ref{fond2})) that

\begin{equation*}
\lambda _{0}=-\alpha \left( \beta _{\pm }\left( \lambda _{2}\right) \pm
\beta _{+}\left( \mu _{2}\right) \right) -2\beta _{\pm }\left( \lambda
_{2}\right) \beta _{+}\left( \mu _{2}\right) .
\end{equation*}

The SUSY partner of $V_{-}(y)$ is:

\begin{eqnarray}
V_{+}(y) &=&V_{-}(y)+2w_{0}^{\prime }(x)=V_{-}(y)+2w_{0}^{\prime }(y)\frac{dy%
}{dx} \\
&=&y^{2}\left( \lambda _{2}\pm 2\alpha \beta _{\pm }\left( \lambda
_{2}\right) \right) +\frac{1}{y^{2}}\left( \mu _{2}+2\alpha \beta _{+}\left(
\mu _{2}\right) \right) +\lambda _{0}+2\alpha \left( \beta _{\pm }\left(
\lambda _{2}\right) \pm \beta _{+}\left( \mu _{2}\right) \right)  \notag
\end{eqnarray}

We define the following multiparameter:

\begin{equation}
a=\left( \lambda ,\mu \right) \left\{ 
\begin{array}{c}
\lambda =\beta _{\pm }\left( \lambda _{2}\right) >0,\ \lambda _{2}=\lambda
\left( \lambda \mp \alpha \right) \\ 
\mu =\beta _{+}\left( \mu _{2}\right) >0,\ \mu _{2}=\mu \left( \mu -\alpha
\right) .%
\end{array}%
\right.
\end{equation}

The initial potential is now

\begin{equation}
V_{-}(y,a)=\lambda \left( \lambda \mp \alpha \right) y^{2}+\frac{\mu \left(
\mu -\alpha \right) }{y^{2}}+\lambda _{0}(a)  \label{potcate2}
\end{equation}%
with 
\begin{equation}
\lambda _{0}(a)=-\alpha (\lambda \pm \mu )-2\lambda \mu  \label{lambda0}
\end{equation}

Its SUSY partner is

\begin{eqnarray}
V_{+}(y,a) &=&\lambda \left( \lambda \pm \alpha \right) y^{2}+\frac{\mu
\left( \mu +\alpha \right) }{y^{2}}+\lambda _{0}(a)+2\alpha \left( \lambda
\pm \mu \right) \\
&=&V_{-}(y,a_{1})+R(a),  \notag
\end{eqnarray}%
where

\begin{equation}
\ \left\{ 
\begin{array}{c}
a_{1}=(\lambda _{1},\mu _{1})=\left( \lambda \pm \alpha ,\mu +\alpha \right)
=a+A_{\pm } \\ 
\ R(a)=\lambda _{0}\left( a\right) -\lambda _{0}\left( a_{1}\right) +2\alpha
\left( \lambda \pm \mu \right) =\pm \left( \phi _{2,\pm }\left( a_{1}\right)
-\phi _{2,\pm }\left( a\right) \right) ,%
\end{array}%
\right.
\end{equation}%
with $A_{\pm }=(\pm \alpha ,\alpha )$\ and: 
\begin{equation}
\phi _{2,\pm }\left( a\right) =\phi _{2,\pm }(\lambda ,\mu )=\left( \lambda
\pm \mu \right) ^{2}.  \label{phi2}
\end{equation}

We recognize (see Eq.(\ref{SIP})) the characteristic behaviour of a TSIP.
More generally

\begin{equation}
\left\{ 
\begin{array}{c}
a_{k}=a+kA=(\lambda _{k},\mu _{k})=\left( \lambda \pm k\alpha ,\mu +k\alpha
\right) \\ 
R(a_{k})=\pm \left( \phi _{2,\pm }\left( a_{k+1}\right) -\phi _{2,\pm
}\left( a_{k}\right) \right) ,%
\end{array}%
\right.
\end{equation}%
The spectrum (see Eq.(\ref{spectreSIP})) becomes

\begin{equation}
E_{n,\pm }(a)=\pm \left( \phi _{2,\pm }\left( a_{n}\right) -\phi _{2,\pm
}\left( a\right) \right)  \label{spectrecat21}
\end{equation}%
or

\begin{equation}
E_{n,\pm }(a)=\pm 4\alpha ^{2}\left( n-\frac{\lambda \pm \mu }{2\alpha }%
\right) ^{2}\mp \phi _{2,\pm }\left( a\right)  \label{spectrecat 21}
\end{equation}

$E_{n,\pm }$ is then an harmonic function of $n+\left( \mu \pm \lambda
\right) /2\alpha $. As for the corresponding superpotential, it takes the
form (see Eq.(\ref{superpot2})):

\begin{equation}
w_{0}(y,a)=\lambda y-\frac{\mu }{y}.  \label{superpotcat21}
\end{equation}

Using Eq.(\ref{RS functions}), Eq.(\ref{spectrecat21}) and Eq.(\ref%
{superpotcat21}), we finally deduce the general form for the RS function
associated to the n$^{th}$ level of the spectrum of an arbitrary potential
of the second category of the considered type:

\begin{eqnarray}
w_{n,\pm }(y,a) &=&\lambda y-\frac{\mu }{y}\mp \frac{\phi _{2,\pm }\left(
a_{n}\right) -\phi _{2,\pm }\left( a\right) }{\left( \lambda +\lambda
_{1}\right) y-\left( \mu +\mu _{1}\right) /y\mp }\Rsh ...
\label{RS functions cat 21} \\
&&  \notag \\
&\Rsh &\frac{\phi _{2,\pm }\left( a_{n}\right) -\phi _{2,\pm }\left(
a_{j-1}\right) }{\left( \lambda _{j-1}+\lambda _{j}\right) y-\left( \mu
_{j-1}+\mu _{j}\right) /y\mp }\Rsh ...  \notag \\
&&  \notag \\
&\Rsh &\frac{\phi _{2,\pm }\left( a_{n}\right) -\phi _{2,\pm }\left(
a_{n-1}\right) }{\left( \lambda _{n-1}+\lambda _{n}\right) y-\left( \mu
_{n-1}+\mu _{n}\right) /y}  \notag
\end{eqnarray}

It is possible to authorize a complex phase $\varphi _{0}$ in the definition
of $y(x)$ (see Eq.(\ref{2 nd category})) wich becomes complex\bigskip . The
real part of corresponding to a shift of the variable, we can suppose $%
\varphi _{0}$ purely imaginary: $\varphi _{0}=i\theta _{0}$. Consider the
case where:

\begin{equation}
\frac{dy}{dx}=\alpha -\alpha y^{2}(x),
\end{equation}

with 
\begin{equation}
y(x)=\tanh \left( \alpha x+i\theta _{0}\right) =-i\frac{\tanh \left( \alpha
x\right) +i\tan \left( \theta _{0}\right) }{\tanh \left( \alpha x\right)
+i\cot \left( \theta _{0}\right) }.
\end{equation}

If $\theta _{0}=\pi /4$, $1/y(x)=\overline{y(x)}$ and the reality of $%
w_{0}(y,a)$ and consequently of every $w_{n,}(y,a)$ is ensured when $\mu =-%
\overline{\lambda }$. We see immediately that $\phi _{2,-}\left( a\right) $,
that is $E_{n,-}(a)$ (cf Eq.(\ref{phi2}) and Eq.(\ref{spectrecat21})), and $%
V_{-}(y,a)$ (cf Eq.(\ref{potcate2})) are also real. We recover in this case
the Scarf II potential (see Annexe B).

Since $\tanh \left( \theta _{0}\right) $ cannot be equal to $1$, the same
reasoning cannot be used when $dy/dx=\alpha +\alpha y^{2}(x)$, that is, when 
$y(x)=\tan \left( \alpha x+\varphi _{0}\right) $. It cannot be neither
adapted for the first category potentials.

\subsection{Comparison with the Barclay-Maxwell classification of TSIP}

\subsubsection{Barclay-Maxwell classification of TSIP}

Concerning the case of TSIP, using exactness arguments of the SWKB condition%
\cite{Cooper,comtet,dutt2}, Barclay and Maxwell\cite{barclay1,Bhaduri} have
established that their associated superpotentials obey one or other of the
following equations:

* Class I potentials:

\begin{equation}
w_{0}^{\prime }(x)=\alpha _{0}+\alpha _{1}w_{0}(x)+\alpha _{2}w_{0}^{2}(x)
\label{potI}
\end{equation}

$\alpha _{i}\in \mathbb{R},$ $i=0,1,2$

* Class II potentials:%
\begin{equation}
w_{0}^{\prime }(x)=\alpha _{0}+\alpha _{1}w_{0}(x)\sqrt{\alpha _{0}+\alpha
_{2}w_{0}^{2}(x)}+\alpha _{2}w_{0}^{2}(x)  \label{potII}
\end{equation}

$\alpha _{i}\in \mathbb{R},$ $i=0,1,2$

Moreover Barclay has shown\cite{barclay2}:

* For Class I potentials, there exists a change of variable $x\rightarrow v$
in which the potential is transformed into that of an harmonic oscillator $%
V(x)\rightarrow V(v)=v^{2}+V_{0}$, $v(x)$ being solution of a Riccati
equation with constant coefficients:

\begin{equation}
\frac{dv(x)}{dx}=A_{0}+A_{1}v(x)+A_{2}v^{2}(x)  \label{potclasse1}
\end{equation}%
with

\begin{equation}
V_{0}=-\frac{\alpha _{1}^{2}}{4\left( 1-\alpha _{2}\right) }-\alpha _{0}.
\label{enerI}
\end{equation}

* For class II potentials, there exists a change of variable $x\rightarrow y$
in which the potential is transformed into that of an harmonic oscillator $%
V(x)\rightarrow V(v)=v^{2}+V_{0}$, $v(x)$ being solution of a first-order
ODE of the form:

\begin{equation}
\frac{dv(x)}{dx}=A_{0}+A_{1}v(x)\sqrt{A_{0}+A_{2}v^{2}(x)}+A_{2}v^{2}(x)
\label{potclasse2}
\end{equation}%
with

\begin{equation}
V_{0}=-\alpha _{0}-\frac{\alpha _{0}\left( 1-\alpha _{2}\right) }{2\alpha
_{2}}\left( 1-\sqrt{1-\frac{\alpha _{1}^{2}\alpha _{2}}{\left( 1-\alpha
_{2}\right) ^{2}}}\right) .  \label{enerII}
\end{equation}

\subsubsection{First category}

\bigskip From Eq.(\ref{potclasse1}) and Eq.(\ref{potreg}), it appears
clearly that the first category of potentials coincides with the above first
class. Eq.(\ref{potI}) is still easily verified using the fact that the
superpotential associated to a first category potential $V(u)=\widetilde{%
\lambda }_{2}u^{2}+\widetilde{\lambda }_{1}u+\widetilde{\lambda }_{0}$ is an
affine function in the $y$ variable (see Eq.(\ref{solpolyedr})) and then
also in the $u$ variable.

Conversely, if $w_{0}(x)$ obeys Eq.(\ref{potI}), then, if we define $%
w_{0}(x)=cu+d,\ c,d\in \mathbb{R}$, $V(x)=-w_{0}^{\prime }(x)+w_{0}^{2}(x)$
becomes quadratic in $u$ and $u(x)$ obeys a constant coefficient Riccati
equation.

\subsubsection{Second category}

For the second category potential $V(y)=\lambda _{2}y^{2}+\lambda _{0}+\mu
_{2}/y^{2}$, with $dy(x)/dx=A_{0}+A_{2}y^{2}(x)$, we have obtained in Eq.(%
\ref{superpotcat21}) that the superpotential is given by:

\begin{equation}
w_{0}=b_{1}y+\frac{b_{-1}}{y}.  \label{superpotential}
\end{equation}

Then:

\begin{equation}
w_{0}^{\prime }(x)=y^{\prime }(x)\left( b_{1}-\frac{b_{-1}}{y^{2}}\right)
=\left( A_{0}b_{1}-A_{2}b_{-1}\right) +A_{2}b_{1}y^{2}-A_{0}b_{-1}\frac{1}{%
y^{2}}.  \label{derivsuperpot2}
\end{equation}

From Eq.(\ref{superpotential}) we have also that

\bigskip 
\begin{equation}
\left\{ 
\begin{array}{c}
y=\frac{1}{2b_{1}}\left( w_{0}\pm \sqrt{w_{0}^{2}-4b_{1}b_{-1}}\right) \\ 
\frac{1}{y}=\frac{1}{2b_{-1}}\left( w_{0}\mp \sqrt{w_{0}^{2}-4b_{1}b_{-1}}%
\right)%
\end{array}%
\right.  \label{y}
\end{equation}%
which combined with Eq.(\ref{derivsuperpot2}) gives

\bigskip 
\begin{equation}
w_{0}^{\prime }(x)=A+Bw_{0}^{2}(x)+C\sqrt{A+Bw_{0}^{2}(x)}
\end{equation}%
with 
\begin{equation}
\left\{ 
\begin{array}{c}
A=2\left( A_{0}b_{1}-A_{2}b_{-1}\right) \\ 
\\ 
B=-A/4b_{1}b_{-1} \\ 
\\ 
C=\pm \left( A_{0}b_{1}+A_{2}b_{-1}\right) /b_{1}b_{-1}\sqrt{B}.%
\end{array}%
\right.
\end{equation}

Consequently any second-category potential is a Class-II potential.

Conversely, suppose $V(x)=-w_{0}^{\prime }(x)+w_{0}^{2}(x)$ is a Class II
potential (see Eq.(\ref{potII})). We then define the $y$ variable via:

\begin{equation}
w_{0}=b_{1}y+\frac{b_{-1}}{y},
\end{equation}%
where $b_{1}b_{-1}=-\alpha _{0}/4\alpha _{2}$. Then using Eq.(\ref{y}) we
have

\begin{equation*}
\sqrt{w_{0}^{2}+\frac{\alpha _{0}}{\alpha _{2}}}=b_{1}y-\frac{b_{-1}}{y}
\end{equation*}%
and from Eq.(\ref{potII}) we deduce that

\begin{eqnarray}
\frac{y^{\prime }}{y}\left( b_{1}y-\frac{b_{-1}}{y}\right) &=&\alpha
_{2}\left( \frac{\alpha _{0}}{\alpha _{2}}+w_{0}^{2}\right) +\alpha _{1}%
\sqrt{\alpha _{2}}w_{0}\sqrt{\frac{\alpha _{0}}{\alpha _{2}}+w_{0}^{2}} \\
&=&\alpha _{2}\left( b_{1}y-\frac{b_{-1}}{y}\right) ^{2}+\alpha _{1}\sqrt{%
\alpha _{2}}\left( b_{1}y+\frac{b_{-1}}{y}\right) \left( b_{1}y-\frac{b_{-1}%
}{y}\right) ,  \notag
\end{eqnarray}%
that is,

\begin{equation}
y^{\prime }=A_{0}+A_{2}y^{2}
\end{equation}%
with

\begin{equation}
\left\{ 
\begin{array}{c}
A_{0}=\left( \alpha _{1}\sqrt{\alpha _{2}}-\alpha _{2}\right) b_{-1} \\ 
A_{2}=\left( \alpha _{1}\sqrt{\alpha _{2}}+\alpha _{2}\right) b_{1}.%
\end{array}%
\right.
\end{equation}

As for the potential, it takes the form

\begin{eqnarray}
V &=&-\alpha _{0}-\alpha _{1}\sqrt{\alpha _{2}}w_{0}\sqrt{\frac{\alpha _{0}}{%
\alpha _{2}}+w_{0}^{2}}+\left( 1-\alpha _{2}\right) w_{0}^{2} \\
&=&-\alpha _{0}-\alpha _{1}\sqrt{\alpha _{2}}\left( b_{1}y+\frac{b_{-1}}{y}%
\right) \left( b_{1}y-\frac{b_{-1}}{y}\right) +\left( 1-\alpha _{2}\right)
\left( b_{1}y+\frac{b_{-1}}{y}\right) ^{2}  \notag \\
&=&\lambda _{2}y^{2}+\lambda _{0}+\frac{\mu _{2}}{y^{2}}  \notag
\end{eqnarray}%
with

\begin{equation}
\left\{ 
\begin{array}{c}
\lambda _{2}=b_{1}^{2}\left( 1-\alpha _{1}\sqrt{\alpha _{2}}-\alpha
_{2}\right) \\ 
\mu _{2}=b_{-1}^{2}\left( 1+\alpha _{1}\sqrt{\alpha _{2}}-\alpha _{2}\right)
\\ 
\lambda _{0}=-\alpha _{0}/2\left( 1+1/\alpha _{2}\right)%
\end{array}%
\right.
\end{equation}

Consequently any Class-II potential is a second-category potential.

Our two categories of potentials coincide then respectively with the two
Barclay-Maxwell classes. \bigskip Since every TSIP is necessarily a member
of either one of these classes\cite{barclay1,Bhaduri}, we can finally
conclude that the rational representations of the excited states of the RS
functions given in Eq.(\ref{RS functions cat1}) and Eq.(\ref{RS functions
cat 21}) are valid for the whole set of TSIP.

\section{Conclusion}

In this paper we have given an exact expression for the RS functions
associated with the excited bound states of every SIP in terms of
terminating continued fractions built on the superpotential (see Eq.(\ref%
{potII})). Considering the set of translationally shape-invariant potentials
(TSIP), we have shown that it can be divided in two categories, coinciding
with the Barclay-Maxwell classes\cite{barclay1,Bhaduri} but using a
different characterization. More precisely any TSIP can be brought into an
harmonic or isotonic form using a change of variable resting on a constant
coefficient Riccati equation without a term of first degree (see Eq.(\ref%
{1st category}) and Eq.(\ref{2 nd category})). It can be noticed that,
according to the Chalykh-Veselov theorem, the harmonic and isotonic
potentials are the only rational isochronous ones and they present
equispaced spectra\cite{Chalykh,asorey}. In terms of this new variable, the
superpotential is a first-degree polynomial or a first-degree Laurent
polynomial (see Eq.(\ref{superpotcat1}) and Eq.(\ref{superpotcat21}))
recovering the Gendenshtein ans\"{a}tze\cite{Gendenshtein}. With this
formulation, the shape invariance characteristics of the potential appear in
a very transparent manner offering a compact expression for the spectrum of
any TSIP (see Eq.(\ref{spectrecat 1}) and Eq.(\ref{spectrecat21})).
Interestingly, we obtain a kind of duality between the TSI potentials and
the associated energy spectrums. For a second category potential which can
be brought into an isotonic form, the energy is an harmonic function of the
shifted quantum number $n\pm a/\alpha $ (see Eq.(\ref{spectrecat1})).
Reciprocally, for a first category potential which can be brought into an
harmonic form, the energy is an isotonic function of $n+\left( \mu \pm
\lambda \right) /2\alpha $ (see Eq.(\ref{spectrecat 21})). Collecting these
results together we obtain compact rational representations (see Eq.(\ref{RS
functions cat1}) and Eq.(\ref{RS functions cat 21}) for all the excited
states RS functions associated to a TSIP.

\section{Acknowledgments}

We would like to thank Professor P.G.L. Leach for useful suggestions and a
careful reading of the manuscript.

\section{Annexe A}

\subparagraph{Rosen-Morse I}

The Rosen-Morse I potential with zero-energy ground state is given by\cite%
{Cooper,Dutt}:

\begin{equation*}
V(x)=\frac{A(A-\alpha )}{\sin ^{2}\left( \alpha x\right) }+\frac{2B}{\tan
\left( \alpha x\right) }+\frac{B^{2}}{A^{2}}-A^{2}.
\end{equation*}

Defining:

\begin{equation}
\left\{ 
\begin{array}{c}
y(x)=\tan \left( \alpha x-\frac{\pi }{2}\right) =-\cot (\alpha x),\ x\in %
\left] 0,\frac{\pi }{\alpha }\right[ \  \\ 
\\ 
y^{\prime }(x)=\alpha +\alpha y^{2}(x),%
\end{array}%
\right.  \label{transfRosenMorse1}
\end{equation}%
we obtain%
\begin{equation}
V(y)=a(a-\alpha )y^{2}+\lambda _{1}y+\frac{\lambda _{1}^{2}}{4a^{2}}-\alpha
a,
\end{equation}%
with $a=A\ $and $\lambda _{1}=-2B$.

The superpotential is:

\begin{equation}
w_{0}(x)=ay+\frac{\lambda _{1}}{2a}=-A\cot (\alpha x)-\frac{B}{A}.
\end{equation}

\subparagraph{Rosen-Morse II}

The Rosen-Morse II potential with zero-energy ground state is given by\cite%
{Cooper,Dutt}:

\subparagraph{%
\protect\begin{equation}
V(y)=-\frac{A(A+\protect\alpha )}{\cosh ^{2}\left( \protect\alpha x\right) }%
+2B\tanh \left( \protect\alpha x\right) +\frac{B^{2}}{A^{2}}+A^{2},\quad
B<A^{2}. 
\protect\end{equation}%
}

Defining:

\begin{equation}
\left\{ 
\begin{array}{c}
y(x)=\tanh \left( \alpha x\right) ,\ x\in \mathbb{R} \\ 
\\ 
y^{\prime }(x)=\alpha -\alpha y^{2}(x),%
\end{array}%
\right.  \label{transfRosenMorse2}
\end{equation}%
we obtain:

\begin{equation}
V(y)=a(a+\alpha )y^{2}+\lambda _{1}y+\frac{\lambda _{1}^{2}}{4a^{2}}-\alpha
a,
\end{equation}%
with $a=A$ and $\lambda _{1}=2B$.

The superpotential is:

\begin{equation}
w_{0}(x)=ay+\frac{\lambda _{1}}{2a}=A\tanh (\alpha x)+\frac{B}{A}.
\end{equation}

\subparagraph{Eckardt}

The Eckardt potential with zero-energy ground state is given by\cite%
{Cooper,Dutt}:

\begin{equation}
V(x)=\frac{A(A-\alpha )}{\sinh ^{2}\left( \alpha x\right) }-2B\coth \left(
\alpha x\right) +\frac{B^{2}}{A^{2}}+A^{2},\quad B>A^{2}.
\end{equation}

Defining:

\begin{equation}
\left\{ 
\begin{array}{c}
y(x)=\coth \left( \left( -\alpha \right) x\right) ,\ x\in \mathbb{R} \\ 
\\ 
y^{\prime }(x)=\left( -\alpha \right) \left( 1-y^{2}(x)\right) ,%
\end{array}%
\right.  \label{transfEckardt}
\end{equation}%
we obtain:

\begin{equation}
V(x)=a(a-\left( -\alpha \right) )y^{2}+\lambda _{1}y+\frac{\lambda _{1}^{2}}{%
4a^{2}}-\left( -\alpha \right) a,
\end{equation}%
with $a=A\ $ and $\lambda _{1}=2B$.

The superpotential is:

\begin{equation}
w_{0}(x)=ay+\frac{\lambda _{1}}{2a}=A\coth \left( \left( -\alpha \right)
x\right) +\frac{B}{A}
\end{equation}

\section{Annexe B}

\subparagraph{P\"{o}schl-Teller}

The P\"{o}schl-Teller potential with zero-energy ground state is given by%
\cite{Cooper,Dutt}:

\begin{equation}
V(x)=A^{2}+\frac{\left( A^{2}+B^{2}+A\alpha \right) }{\sinh ^{2}\left(
\alpha x\right) }-B(2A+\alpha )\frac{\coth \left( \alpha x\right) }{\sinh
\left( \alpha x\right) },\quad B>A,\quad x>0
\end{equation}

Defining:

\begin{equation}
\left\{ 
\begin{array}{c}
y(x)=\tanh \left( \frac{\alpha x}{2}\right) \\ 
\\ 
y^{\prime }(x)=\frac{\alpha }{2}-\frac{\alpha }{2}y^{2}(x),%
\end{array}%
\right.  \label{transfPT}
\end{equation}%
we obtain:

\begin{equation}
V(y)=\lambda \left( \lambda +\alpha /2\right) y^{2}+\frac{\mu \left( \mu
-\alpha /2\right) }{y^{2}}-\frac{\alpha }{2}(\lambda -\mu )-2\lambda \mu ,
\end{equation}%
with $\lambda =\left( A+B\right) /2$ and $\mu =\left( B-A\right) /2$.

The superpotential is:

\begin{equation}
w_{0}(x)=\lambda y-\frac{\mu }{y}=\frac{A+B}{2}\ \tanh (\frac{\alpha x}{2})-%
\frac{\left( B-A\right) /2}{\tanh (\frac{\alpha x}{2})}=\frac{A}{\tanh
(\alpha x)}-\frac{B}{\sinh (\alpha x)}.
\end{equation}

\subparagraph{P\"{o}schl-Teller I}

The P\"{o}schl-Teller I potential with zero-energy ground state is given by%
\cite{Cooper,Dutt}:

\begin{equation}
V(x)=-(A+B)^{2}+\frac{A(A-\alpha )}{\cos ^{2}\left( \alpha x\right) }+\frac{%
B(B-\alpha )}{\sin ^{2}\left( \alpha x\right) },\quad x\in \left] 0,\frac{%
\pi }{2\alpha }\right[ \ 
\end{equation}

Defining:

\begin{equation}
\left\{ 
\begin{array}{c}
y(x)=\tan \left( \alpha x\right) \  \\ 
\\ 
y^{\prime }(x)=\alpha +\alpha y^{2}(x).%
\end{array}%
\right.  \label{transfPT1}
\end{equation}%
we obtain:

\begin{equation}
V(y)=\lambda \left( \lambda -\alpha \right) y^{2}+\frac{\mu \left( \mu
-\alpha \right) }{y^{2}}-\alpha (\lambda +\mu )-2\lambda \mu ,
\end{equation}%
with $\lambda =A\ $\ and $\mu =B$.

The superpotential is:

\begin{equation}
w_{0}(x)=\lambda y-\frac{\mu }{y}=A\tan \left( \alpha x\right) \ -\frac{B}{%
\tan (\alpha x)}.
\end{equation}

\subparagraph{P\"{o}schl-Teller II}

The P\"{o}schl-Teller II potential with zero-energy ground state is given by%
\cite{Cooper,Dutt}:

\begin{equation}
V(x)=(A-B)^{2}-\frac{A(A+\alpha )}{\cosh ^{2}\left( \alpha x\right) }+\frac{%
B(B-\alpha )}{\sinh ^{2}\left( \alpha x\right) },\quad B<A\ ,\quad x>0
\end{equation}

\bigskip Defining:

\begin{equation}
\left\{ 
\begin{array}{c}
y(x)=\tanh \left( \alpha x\right) <1 \\ 
\\ 
y^{\prime }(x)=\alpha -\alpha y^{2}(x).%
\end{array}%
\right.  \label{transfPT2}
\end{equation}%
we obtain:%
\begin{equation}
V(y)=\lambda \left( \lambda +\alpha \right) y^{2}+\frac{\mu \left( \mu
-\alpha \right) }{y^{2}}-\alpha (\lambda -\mu )-2\lambda \mu ,
\end{equation}%
with

\begin{equation}
\left\{ 
\begin{array}{c}
\lambda =A\  \\ 
\\ 
\mu =B\ .%
\end{array}%
\right.
\end{equation}

The superpotential is:

\begin{equation}
w_{0}(y)=\lambda y-\frac{\mu }{y}=A\tanh \left( \alpha x\right) \ -\frac{B}{%
\tanh (\alpha x)}.
\end{equation}

\subparagraph{Scarf I}

The Scarf I potential with zero-energy ground state is given by\cite%
{Cooper,Dutt}:

\begin{equation}
V(x)=-A^{2}+\frac{\left( A^{2}+B^{2}-A\alpha \right) }{\sin ^{2}\left(
\alpha x\right) }-B(2A-\alpha )\frac{\cot \left( \alpha x\right) }{\sin
\left( \alpha x\right) },\quad B<A,\quad ,\quad x\in \left] 0,\frac{\pi }{%
2\alpha }\right[
\end{equation}

Defining:

\begin{equation}
\left\{ 
\begin{array}{c}
y(x)=\tan \left( \frac{\alpha x}{2}\right) \\ 
\\ 
y^{\prime }(x)=\frac{\alpha }{2}+\frac{\alpha }{2}y^{2}(x).%
\end{array}%
\right.  \label{transscarf1}
\end{equation}%
we obtain:%
\begin{equation}
V(y)=\lambda \left( \lambda -\alpha /2\right) y^{2}+\frac{\mu \left( \mu
-\alpha /2\right) }{y^{2}}-\frac{\alpha }{2}(\lambda +\mu )-2\lambda \mu ,
\end{equation}%
with $\lambda =\left( A+B\right) /2\ $ and $\mu =\left( A-B\right) /2\ $.

The superpotential is:

\begin{equation}
w_{0}(x)=\lambda y-\frac{\mu }{y}=\frac{A+B}{2}\ \tan (\frac{\alpha x}{2})-%
\frac{\left( A-B\right) /2}{\tan (\frac{\alpha x}{2})}=-\frac{A}{\tan
(\alpha x)}+\frac{B}{\sin (\alpha x)}.
\end{equation}

\subparagraph{Scarf II}

The Scarf II potential with zero-energy ground state is given by\cite%
{Cooper,Dutt}:

\begin{equation}
V(x)=A^{2}+\frac{\left( B^{2}-A^{2}-A\alpha \right) }{\cosh ^{2}\left(
\alpha x\right) }-B(2A+\alpha )\frac{\tanh \left( \alpha x\right) }{\cosh
\left( \alpha x\right) }.
\end{equation}

Defining:

\begin{equation}
\left\{ 
\begin{array}{c}
y(x)=\tanh \left( \frac{\alpha x}{2}+i\frac{\pi }{4}\right) \\ 
\\ 
y^{\prime }(x)=\frac{\alpha }{2}-\frac{\alpha }{2}y^{2}(x).%
\end{array}%
\right.
\end{equation}%
we obtain:%
\begin{equation}
V(y)=\lambda \left( \lambda +\alpha /2\right) y^{2}+\frac{\mu \left( \mu
-\alpha /2\right) }{y^{2}}-\frac{\alpha }{2}(\lambda -\mu )-2\lambda \mu ,
\end{equation}%
with $\lambda =\left( A+iB\right) /2$ and $\mu =\left( -A+iB\right) /2$.

The superpotential is:

\begin{equation}
w_{0}(x)=\lambda y-\frac{\mu }{y}=\frac{A+iB}{2}\ \tanh \left( \frac{\alpha x%
}{2}+i\frac{\pi }{4}\right) -\frac{\left( -A+iB\right) /2}{\tanh \left( 
\frac{\alpha x}{2}+i\frac{\pi }{4}\right) }=A\tanh (\alpha x)+\frac{B}{\cosh
(\alpha x)}.
\end{equation}



\begin{thebibliography}{99}
\bibitem{Carinena} J.\ F.\ Cari\~{n}ena, A.\ Ramos and D.\ J.\ Fernandez,
\textquotedblleft Group theoretical approach to the intertwined
hamiltonians,\textquotedblright\ Ann. Phys., \textbf{292}, 42-66 (2001).

\bibitem{Ramos} J.\ F.\ Cari\~{n}ena and A.\ Ramos, \textquotedblleft
Integrability of Riccati equation from a group theoretical
viewpoint,\textquotedblright\ Int. J. Mod. Phys. A, \textbf{14}, 1935-1951
(1999).

\bibitem{Fernandez} D.\ J.\ Fernandez, V.\ Hussin and B.\ Mielnik,
\textquotedblleft A simple generation of exactly solvable anharmonic
oscillators,\textquotedblright\ Phys. Lett. A, \textbf{244} 309-316\ (1998).

\bibitem{Mielnik} B. Mielnik, L.M. Nieto and O. Rosas--Ortiz,
\textquotedblleft The finite difference algorithm for higher order
supersymmetry,\textquotedblright\ Phys. Lett. A, \textbf{269} 70-78\ (2000).

\bibitem{Adler1} V. E. Adler, \textquotedblleft Recuttings of
polygons,\textquotedblright\ Funct. Anal. and Its Appl., \textbf{27} 141-143
(1993).

\bibitem{Adler2} V. E. Adler, \textquotedblleft Nonlinear chains and Painlev%
\'{e} equations,\textquotedblright\ Physica D, \textbf{73} 335-351 (1994).

\bibitem{Cooper} F.\ Cooper, A.\ Khare and U.\ Sukhatme, \textit{%
Supersymmetry in Quantum Mechanics} (World Scientific, Singapore, 2001).

\bibitem{Dutt} R.\ Dutt, A. Khare and U.\ P.\ Sukhatme, \textquotedblleft
Supersymmetry, shape invariance and exactly solvable
potentials,\textquotedblright\ Am. J. Phys. 5\textbf{6}, 163--168 (1988).

\bibitem{Gendenshtein} L.\ Gendenshtein, \textquotedblleft Derivation of
exact spectra of the Schrodinger equation by means of
supersymmetry,\textquotedblright\ JETP Lett. \textbf{38}, 356-359 (1983).

\bibitem{Carinena2} J.\ F.\ Cari\~{n}ena and A.\ Ramos, \textquotedblleft
Riccati equation, factorization method and shape
invariance,\textquotedblright\ Rev. Math. Phys, \textbf{12}, 1279-1304
(2000).

\bibitem{Kazimierz} R.\ Kazimierz, \textquotedblleft The shape invariance
condition,\textquotedblright\ Concepts of Phys., vol 3,
merlin.fic.uni.lodz.pl/concepts/2006.

\bibitem{barclay1} D. T. Barclay and C.J. Maxwell, \textquotedblleft Shape
invariance and the SWKB series,\textquotedblright\ Phys. Lett. A \textbf{157}%
, 357-360 (1991).

\bibitem{Bhaduri} R.K. Bhaduri, J.\ Sakhr, D.W.L.\ Sprung, R.\ Dutt and A.\
Suzuki, \textquotedblleft Shape invariant potentials in SUSY quantum
mechanics and periodic orbit theory,\textquotedblright\ J. Phys. A, \textbf{%
38}, L183 (2005).

\bibitem{bhalla} R.\ S.\ Bhalla, A.\ K.\ Kapoor and P.\ K.\ Panigrahi,
\textquotedblleft Exactness of the supersymmetric WKB approximation
scheme,\textquotedblright\ Phys. Rev. A, \textbf{54}, 951-954 (1996).

\bibitem{bhalla2} R.\ S.\ Bhalla, A.\ K.\ Kapoor and P.\ K.\ Panigrahi,
\textquotedblleft Energy Eigenvalues for a class of one-dimensional
potentials via Quantum Hamilton-Jacobi formalism,\textquotedblright\ Mod.
Phys. Lett. A, \textbf{12}, 295-306 (1997).

\bibitem{rasinariu} C.\ Rasinariu, J.\ J.\ Dykla, A.\ Gangopadhyaya and J.\
V.\ Mallow, \textquotedblleft Exactly solvable systems and the quantum
Hamilton-Jacobi formalism,\textquotedblright\ Phys Lett. A \textbf{338},
197-202 (2005).

\bibitem{weissman} Y. Weissman and J. Jortner, \textquotedblleft The
isotonic oscillator\textquotedblright , Phys. Lett. A 70, 177--179 (1979).

\bibitem{zhu} D. Zhu, \textquotedblleft A new potential with the spectrum of
an isotonic oscillator\textquotedblright\ J. Phys. A 20,4331--4336 (1987).

\bibitem{comtet} A.\ Comtet, A.D.\ Bandrauk abd D.K.\ Campbell,
\textquotedblleft Exactness of semiclassical bound state energies for
supersymmetric quantum mechanics,\textquotedblright\ Phys. Lett. B \textbf{%
150}, 159-162 (1985).

\bibitem{dutt2} R.\ Dutt, A. Khare and U.\ P.\ Sukhatme, \textquotedblleft
Supersymmetry-inspired WKB approximation in quantum
mechanics,\textquotedblright\ Am. J. Phys. 5\textbf{9}, 723--727 (1991).

\bibitem{barclay2} D. T. Barclay \textquotedblleft Convergent WKB
series,\textquotedblright\ Phys Lett. A \textbf{185}, 169-173 (1994).

\bibitem{Chalykh} O.\ A.\ Chalykh and A.\ P.\ Veselov, \textquotedblleft A
remark on rational isochronous potentials,\textquotedblright\ J.\ Nonlin.
Math. Phys. \textbf{12,} Supp. 1, 179-183 (2005).

\bibitem{asorey} M.\ Asorey, J.\ F.\ Cari\~{n}ena, G.\ Marmo and A.\
Perelomov \textquotedblleft Isoperiodic classical systems and their quantum
counterparts,\textquotedblright\ Ann. Phys., \textbf{322}, 1444-1465 (2007)
\end{thebibliography}
\end{document}